\documentclass[twocolumn,aps,prd,preprintnumbers,showpacs,superscriptaddress,nofootinbib,amsmath,amssymb,floats,floatfix,showkeys,notitlepage,longbibliography]{revtex4-1}
\usepackage{comment}
\usepackage{graphicx}
\usepackage{subfigure}
\usepackage{palatino}
\usepackage[commandnameprefix=always]{changes}
\usepackage{hyperref}
\hypersetup{colorlinks=true,linkcolor=blue,urlcolor=blue,citecolor=blue}
\usepackage[toc,page]{appendix}
\usepackage[normalem]{ulem}

\usepackage{orcidlink}
\usepackage{lipsum}
\usepackage{graphicx}
\usepackage{subfigure}
\usepackage{palatino}
\usepackage{sans}
\usepackage{adjustbox}
\usepackage{latexsym}
\usepackage{amsmath}
\usepackage{amssymb}
\usepackage{amsfonts}
\usepackage{dcolumn}
\usepackage{bm}
\usepackage{tikz}
\usepackage{bigints}
\usepackage{array,tabularx,multirow,booktabs}
\usepackage[tracking=true]{microtype}
\SetTracking{}{500}
\SetTracking{encoding={*}, shape=sc}{40}
\UseRawInputEncoding 
\allowdisplaybreaks
\usepackage{adjustbox}
\usepackage{latexsym}
\usepackage{amsmath}
\usepackage{amssymb}
\usepackage{amsfonts}
\usepackage{dcolumn}
\usepackage{bm}
\usepackage{tikz}
\usepackage{bigints}
\usepackage{array,tabularx,multirow,booktabs}
\usepackage[tracking=true]{microtype}
\usepackage{color}

\def\1o2{{1\over2}}

\UseRawInputEncoding 
\allowdisplaybreaks

\begin{document} 

\title{Shadow of the generalized Vaidya black hole}

\author{V. Vertogradov
\orcidlink{0000-0002-5096-7696}}
\email{vdvertogradov@gmail.com}
\affiliation{Wilczek Quantum Center, Shanghai Institute for Advanced Studies, Shanghai, 201315, China.}
\affiliation{University of Science and Technology of China, Hefei, 230026, China}
\affiliation{Physics department, Herzen state Pedagogical University of Russia,
48 Moika Emb., Saint Petersburg 191186, Russia.}
\affiliation{Center for Theoretical Physics, Khazar University, 41 Mehseti Street, Baku, AZ-1096, Azerbaijan.}
\affiliation{SPB branch of SAO RAS, 65 Pulkovskoe Rd, Saint Petersburg
196140, Russia.}

\author{A. \"Ovg\"un
\orcidlink{0000-0002-9889-342X}
}
\email{ali.ovgun@emu.edu.tr}
\affiliation{Physics Department, Eastern Mediterranean
University, Famagusta, 99628 North Cyprus, via Mersin 10, Turkiye.}

\begin{abstract}
In this paper, we investigate the shadow of the generalized Vaidya black hole described by the Husain solution. 
Assuming a barotropic equation of state for the additional matter sector, we first identify the self-similar class admitting a homothetic Killing vector and transform the geometry to a conformally static form. 
This allows us to derive the effective potential for null geodesics in the conformally static representative, determine the corresponding homothetic photon surface, and then transform the result back to the original generalized Vaidya coordinates. We show that, within the weak-energy-condition branches, the parameter $\alpha$ controlling the equation of state determines whether the shadow is enlarged or reduced relative to the Vaidya case: the branch $0\leq\alpha<1/2$ increases the photon-sphere and shadow radii, whereas the branch $1/2<\alpha\leq 1$ decreases them. We then turn to the genuinely time-dependent Husain metric and derive 
\textcolor{black}{quasistatic} slow-evolution equations for the instantaneous photon surface and 
critical impact parameter.
In particular, we show that the growth or contraction of the shadow is governed by the local effective influx in the photon region rather than by the signs of $\dot M$ and $\dot N$ separately. 
For the charge-like branch, this yields a transparent criterion for when accretion enlarges the shadow and when rapid charge growth instead causes it to shrink.
\end{abstract}

\date{\today}



\maketitle

\section{Introduction}

Black hole shadows probe the geometry of null geodesics in the strongest accessible gravitational fields. In static and stationary spacetimes, the shadow boundary is determined by unstable photon orbits and is therefore directly tied to photon spheres, photon regions, and critical impact parameters. This close relation between optics and geometry makes shadow observables a particularly useful tool for testing black hole spacetimes, constraining deviations from Kerr, and connecting horizon-scale imaging to the analytical structure of null geodesics~\cite{1988GReGr..20.1173A,Claudel:2000yi,Cunha:2018acu,bib:tsupko_review,Gralla:2019xty}. This connection has motivated a vast literature on black hole shadows in vacuum, matter-dressed, and modified-gravity spacetimes, including rotating non-Kerr geometries, regular black holes, black holes in plasma environments, and quantum-corrected models~\cite{Atamurotov:2013sca,Abdujabbarov:2015xqa,Younsi:2016azx,Abdujabbarov:2016hnw,Abdujabbarov:2016efm,Wang:2017hjl,Tsukamoto:2017fxq,Konoplya:2019sns,Konoplya:2020bxa,Kumar:2020owy,Afrin:2021imp,Atamurotov:2021hoq,Okyay:2021nnh,Pantig:2022ely,Pantig:2022gih,Pantig:2022qak,Kuang:2022xjp,Kuang:2022ojj,Meng:2022kjs,Pulice:2023dqw,Ovgun:2024zmt,Lambiase:2024uzy,Lambiase:2024vkz,Karshiboev:2024xxx,Liu:2024soc,Liu:2024lbi,Heidari:2024bkm}.

The observational importance of this subject was dramatically strengthened by the Event Horizon Telescope collaboration, which reported horizon-scale images of M87$^*$ in 2019 and Sagittarius A$^*$ in 2022. 
These observations provided direct evidence for compact dark central regions surrounded by bright lensed emission rings and opened a new window for testing strong-gravity models through horizon-scale imaging~\cite{bib:m87,EventHorizonTelescope:2022wkp,bib:vanozi:2022moj}. 
At the same time, it has become increasingly clear that extracting spacetime information from black hole images requires a precise understanding of the relation between shadows, photon rings, lensing rings, and the astrophysical environment~\cite{bib:falke,Gralla:2019xty,Kumar:2018ple,Adler:2022qtb}. 
For this reason, analytical control of the photon-sphere/shadow correspondence remains essential.

Most analytical studies have focused on static or stationary spacetimes, where Killing symmetries provide conserved quantities that simplify the null-geodesic problem. 
By contrast, genuinely dynamical black holes are considerably harder to analyze. 
In a time-dependent geometry one generally loses the timelike Killing vector, so the usual notion of conserved photon energy is absent and the standard effective-potential approach must be reformulated. 
This difficulty is not merely technical: accretion, evaporation, collapse, and interactions with surrounding matter all naturally introduce time dependence, so realistic black hole environments are not expected to be exactly stationary. 
Understanding shadows in dynamical spacetimes is therefore an important step toward bridging the gap between idealized black hole optics and physically evolving compact objects~\cite{bib:tsupko_review,bib:understanding,bib:perlik2022prd,Tan:2023ngk,Wang:2025fmz}.  

A particularly useful setting for this problem is provided by the Vaidya family of solutions. 
The original Vaidya metric describes a spherically symmetric spacetime sourced by null radiation and has long played a central role in studies of radiating stars, accreting compact objects, gravitational collapse, dynamical horizons, and Hawking-like processes~\cite{bib:vaidya,bib:124,bib:124L,bib:nelvaidya,bib:nelsurface,bib:diag,bib:charged_hak}. 
Because of its simplicity, the Vaidya geometry has also become one of the standard laboratories for dynamical black hole optics. 
The first analytical studies of shadows in nonstationary black hole backgrounds were carried out in cosmological and conformally related settings~\cite{bib:tsupko_first,PerlickEtAl:2018,SchneiderPerlick:2018}, while later work clarified the evolution of photon spheres and shadows in dynamical spacetimes more generally~\cite{bib:understanding,bib:japan}. 
A major advance was achieved for self-similar Vaidya black holes, where a homothetic symmetry allows one to map the problem to a conformally static geometry and derive exact analytical formulas for the photon sphere and angular shadow radius~\cite{bib:perlik2022prd}.  

These developments naturally motivate extending the analysis beyond pure null radiation. 
The generalized Vaidya spacetime introduces an additional matter sector on top of the null-fluid contribution and provides a much richer class of dynamical spherical geometries~\cite{bib:vunk}. 
Among its most important exact realizations is the Husain solution, obtained when the extra matter sector satisfies a barotropic equation of state \(P=\alpha\rho\)~\cite{bib:husain1996}. 
This solution and its generalizations have been widely used in studies of gravitational collapse, relativistic stellar atmospheres, conformal symmetries, naked singularities, horizon structure, and generalized radiating configurations~\cite{bib:Radiationstring1998,bib:twofluidatm1999,bib:Maombi1,bib:ver1,bib:ver2,bib:ver3,bib:myrev,bib:r2,bib:maharaj,bib:vertogradov2023mpla,bib:vertogradov2024grg,bib:vertogradov2022universe,bib:vertogradov2020universe,bib:yaghoub2017epjc,bib:yaghoub2017epjc2,bib:yaghoub2018epjc}. 
From the optical point of view, the Husain spacetime is especially attractive because, for a suitable self-similar choice of the free functions, it also admits a homothetic Killing vector. 
This makes it possible to treat a nontrivial matter-dressed dynamical black hole with analytical methods closely analogous to those used in the self-similar Vaidya case.

There is a second, more physical, reason to study this geometry. 
Recent investigations of regular and nonsingular black hole formation suggest that during collapse the matter sector may pass through phases with nontrivial effective equations of state, while the observable exterior geometry remains of generalized Vaidya/Husain type. 
In such situations, shadow observables are controlled primarily by the external optical geometry rather than by the detailed microphysics of the deep interior. 
This makes the Husain solution a natural effective exterior model for testing how a barotropic matter sector modifies photon capture and the apparent size of the black hole shadow~\cite{bib:ali2024podu,bib:ali2024plb,bib:vertogradov2023pocs,bib:vertogradov2024epjplus,bib:yaghoub2024epjc}. 

The central question we address in this paper is therefore the following: 
\emph{how does the barotropic matter sector in the generalized Vaidya/Husain spacetime alter the photon sphere and the shadow relative to the self-similar Vaidya background, and how do these optical characteristics evolve in the genuinely dynamical case?}
This question is timely for at least three reasons. 
First, it extends the analytical theory of dynamical shadows from pure Vaidya radiation to a broader and physically richer matter source. 
Second, the parameter \(\alpha\) directly controls the sign and radial scaling of the additional matter contribution, so it provides a clean handle on how the equation of state influences null geodesics and optical observables. 
Third, in the branch \(\alpha>1/2\), the additional term acquires a charge-like character, allowing one to discuss dynamical processes such as charged accretion and neutralization in terms of shadow evolution.

The main aim of the present work is to develop an analytical framework for the shadow of the Husain black hole in both the self-similar and slowly evolving regimes. 
More specifically, we first identify the self-similar subclass for which the metric admits a homothetic Killing vector and rewrite the spacetime in conformally static coordinates. 
This enables us to derive the null-geodesic equations, the effective potential, and the conditions for unstable circular photon orbits. 
We then compare the resulting photon sphere and shadow with those of the self-similar Vaidya solution and show that the two weak-energy-condition branches, \(0\le \alpha<1/2\) and \(1/2<\alpha\le 1\), lead to qualitatively different optical behavior. 
In addition, we present exact benchmark cases for \(\alpha=0\) and \(\alpha=1\), which provide nonperturbative support for the general branch structure. 
Finally, we turn to the genuinely dynamical Husain geometry and derive local \textcolor{black}{quasistatic} criteria governing the evolution of the photon sphere and the shadow in terms of the effective influx in the photon region. 
This formulation clarifies when accretion enlarges the shadow and when the growth of the charge-like sector causes it instead to shrink.

The paper is organized as follows. 
In Sec.~II we review the generalized Vaidya/Husain solution and the associated matter content. 
In Sec.~III we isolate the self-similar branch admitting a homothetic Killing vector and transform the metric to conformally static form. 
In Sec.~IV we revisit the self-similar Vaidya shadow as the reference case and then analyze the photon sphere and shadow of the Husain spacetime, including exact benchmark branches. 
In Sec.~V we investigate the \textcolor{black}{quasistatic} slow evolution of the photon surface 
and shadow in the general time-dependent Husain geometry. 	Finally, in Sec.~VI we summarize the results and discuss their physical implications.

\section{Husain solution}

We consider the generalized Vaidya spacetime described by the metric \cite{bib:husain1996}
\begin{equation}
ds^2 = -\left(1-\frac{2M(v,r)}{r}\right)dv^2+2dvdr+r^2d\Omega^2,
\end{equation}
where
\begin{equation}
d\Omega^2=d\theta^2+\sin^2\theta\,d\varphi^2
\end{equation}
is the metric on the unit two-sphere. The generalized Vaidya geometry is an exact solution of the Einstein equations sourced by two different matter contributions: a type-II null radiation component, which reduces to the standard Vaidya source \cite{bib:vaidya}, and a type-I matter component describing an additional matter distribution. The total energy-momentum tensor can be written as
\begin{equation}
T_{ik}=T^{(n)}_{ik}+T^{(m)}_{ik},
\end{equation}
where $T^{(n)}_{ik}$ denotes the null radiation part and $T^{(m)}_{ik}$ denotes the type-I matter contribution.

The corresponding components are
\begin{align}
T^{(n)}_{ik}&=\mu_{(n)}L_iL_k.\\
T^{(m)}_{ik} &= (\rho+P)(L_iN_k+L_kN_i)+Pg_{ik},
\end{align}
where
\begin{align}
\mu_{(n)} = \frac{2\dot{M}}{r^2},\\
\rho &= \frac{2M'}{r^2},\\
P &= -\frac{M''}{r},\\
L_i &= \delta^0_i,\\
N_i &= \frac{1}{2}\left(1-\frac{2M}{r}\right)\delta^0_i-\delta^1_i.
\end{align}
Here a dot denotes differentiation with respect to $v$, whereas a prime denotes differentiation with respect to $r$. The null vectors satisfy
\begin{align}
L_iL^i &= N_iN^i=0,\\
L_iN^i &= -1.
\end{align}
Thus, $\mu$ is the energy density of the null radiation, while $\rho$ and $P$ are the energy density and pressure of the type-I matter sector, respectively.

The physical viability of the matter source is constrained by the standard energy conditions. 
For the type-II null-radiation sector, the null and weak energy conditions require
\begin{equation}
\mu_{(n)}\geq 0.
\end{equation}
For the type-I matter sector with energy density $\rho$ and isotropic pressure $P$, the weak and null energy conditions require
\begin{equation}
\rho\geq 0,
\qquad
\rho+P\geq 0.
\end{equation}
If one further imposes the dominant energy condition, one obtains
\begin{equation}
\rho\geq |P|.
\end{equation}
For the barotropic equation of state
\begin{equation}
P=\alpha \rho,
\end{equation}
the dominant energy condition is therefore equivalent to
\begin{equation}
|\alpha|\leq 1,
\end{equation}
provided $\rho\geq 0$. In the present work, we restrict attention to the physically most relevant range
\begin{equation}
0\leq \alpha \leq 1,
\qquad
\alpha\neq \frac12,
\end{equation}
for which the pressure is non-negative and the standard energy conditions are mutually compatible.

Solving the Einstein equations with the barotropic equation of state $P=\alpha\rho$, one finds that the mass function takes the form

\begin{equation}
M(v,r)=M(v)+\frac{1}{2}N(v)r^{1-2\alpha},
\end{equation}
where $\alpha\neq 1/2$ and $\alpha\in[-1,1]$. The function $N(v)$ characterizes the strength of the additional matter sector. In particular, in the charged interpretation supported by nonlinear electrodynamics, the parameter $N$ represents a combination of magnetic and electric charges rather than a purely Maxwellian charge parameter~\cite{bib:vertogradov2026ijgmmp}. This point is important because the corresponding charge contribution is generated by a nonlinear electromagnetic source.

For later use, it is convenient to record the corresponding matter variables explicitly:
\begin{equation}
\rho=\frac{(1-2\alpha)N(v)}{r^{2\alpha+2}},
\qquad
P=\alpha \rho.
\end{equation}
Hence the condition $\rho\geq 0$ implies
\begin{align}
N(v)>0, \qquad & 0\leq \alpha < \frac12,\\
N(v)<0, \qquad & \frac12 < \alpha \leq 1.
\end{align}
Therefore, the sign of the additional Husain contribution is fixed by the equation-of-state parameter once the weak energy condition is imposed.

Substitution of this mass function into the metric gives
\begin{equation}
ds^2=-\left(1-\frac{2M(v)}{r}-\frac{N(v)}{r^{2\alpha}}\right)dv^2+2dvdr+r^2d\Omega^2.
\end{equation}

If the spacetime admits a conformal Killing vector, the functions $M(v)$ and $N(v)$ acquire the specific form
\begin{align}
N(v) &= \nu v^{2\alpha},\\
M(v) &= \mu v.
\end{align}
Here the constant $\mu$ entering $M(v)=\mu v$ should not be confused with the radiation density denoted by the same symbol above. With this choice, the metric becomes
\begin{equation}
ds^2=-\left(1-\frac{2\mu v}{r}-\frac{\nu v^{2\alpha}}{r^{2\alpha}}\right)dv^2+2dvdr+r^2d\Omega^2.
\end{equation}

We now perform the coordinate transformation
\begin{align} \label{eq:transform1234}
v &= r_0e^{\frac{t}{r_0}},\\
r &= Re^{\frac{t}{r_0}}.
\end{align}
Then the metric becomes conformally static,
\begin{equation}
ds^2=e^{2\frac{t}{r_0}}ds_{cs}^2,
\end{equation}
where
\begin{equation} \label{eq:conformstatic}
ds_{cs}^2=-\left(1-\frac{2\mu r_0}{R}-\frac{\nu(\alpha)r_0^{2\alpha}}{R^{2\alpha}}-\frac{2R}{r_0}\right)dt^2+2dtdR+R^2d\Omega^2.
\end{equation}
\textcolor{black}{It is important to clarify the physical meaning of the coordinates used above.
The transformation \eqref{eq:transform1234} is time dependent when expressed in terms of the original advanced coordinate $v$. Its inverse form is}
\begin{equation} \label{eq:transform4321}
t=r_0\ln\frac{v}{r_0},
\qquad
R=\frac{r_0 r}{v}.
\end{equation}
\textcolor{black}{Consequently, a surface of constant $R$ is not a surface of constant areal radius in the original generalized Vaidya spacetime. Instead,}
\begin{equation}
R=R_*
\qquad
\Longleftrightarrow
\qquad
r_*(v)=\frac{R_*}{r_0}v.
\end{equation}
\textcolor{black}{Thus constant $R$ curves are homothetic curves adapted to the self-similar structure of the spacetime.}

\textcolor{black}{This observation is essential for the interpretation of the optical quantities computed below. Since null geodesics are invariant under conformal transformations up to reparametrization, the conformally static metric $ds_{cs}^2$ is a useful auxiliary metric for deriving null trajectories. However, a circular photon orbit at constant $R=R_{\rm ph}$ in the conformally static coordinates corresponds in the original coordinates to the self-similar photon surface}
\begin{equation}
r_{\rm ph}(v)=\frac{R_{\rm ph}}{r_0}v.
\end{equation}
\textcolor{black}{Therefore, throughout the following discussion, $R_{\rm ph}$ should be understood as the radius of a homothetic photon surface in the conformally static coordinates, not as the areal radius of an ordinary static photon sphere in the original dynamical spacetime.}

\textcolor{black}{Similarly, an observer located at fixed $R=R_o$ corresponds in the original coordinates to the homothetic worldline}
\begin{equation}
r_o(v)=\frac{R_o}{r_0}v.
\end{equation}
\textcolor{black}{The shadow angle derived below is therefore the angular size measured by this class of homothetic observers.}

The energy conditions impose a sign restriction on the parameter $\nu(\alpha)$:
\begin{align}
\nu(\alpha) &= \nu>0,\qquad \alpha<\frac{1}{2},\\
\nu(\alpha) &= -\nu<0,\qquad \alpha>\frac{1}{2}.
\end{align}
Therefore, the same formal term in the metric has different physical signs depending on the equation-of-state parameter $\alpha$.

The conformally static coordinates admit the Killing vector $\partial_t$. This Killing symmetry is the conformal counterpart of the homothetic symmetry of the original self-similar Husain spacetime. Therefore, for null geodesics, one may use the conserved quantity associated with $\partial_t$ in the conformally related metric. This is a useful tool for constructing the null trajectories; the resulting optical quantities must then be mapped back to the original coordinates by means of \eqref{eq:transform4321}. Energy  and angular momentum can be written for conformally-static metric in the form:

\begin{align}
E &= \left(1-\frac{2\mu r_0}{R}-\frac{\nu r_0^{2\alpha}}{R^{2\alpha}}-\frac{2R}{r_0}\right)\frac{dt}{d\lambda}-\frac{dR}{d\lambda},\\
L &= R^2\frac{d\varphi}{d\lambda}.
\end{align}
To study the shadow, we consider null geodesics satisfying
\begin{equation}
g_{ik}u^iu^k=0.
\end{equation}
Due to spherical symmetry, the motion can be restricted to the equatorial plane,
\begin{equation}
\theta=\frac{\pi}{2}.
\end{equation}
The radial equation can then be written as
\begin{equation}
\left(\frac{dR}{d\lambda}\right)^2+V_{\text{eff}}=0,
\end{equation}
with the effective potential
\begin{equation}\label{eq:potential}
V_{\text{eff}}=
\left(1-\frac{2\mu r_0}{R}-\frac{\nu r_0^{2\alpha}}{R^{2\alpha}}-\frac{2R}{r_0}\right)\frac{L^2}{R^2}-E^2.
\end{equation}

The homothetic photon surface is determined by the conditions

\begin{equation}
V_{\text{eff}}=0,\qquad \frac{dV_{\text{eff}}}{dR}=0.
\end{equation}
Equivalently, introducing the impact parameter
\begin{equation}
b\equiv\frac{L}{E},
\end{equation}
one obtains
\begin{align}\label{eq:conditions}
f'(R_{\text{ph}})R_{\text{ph}}-2f(R_{\text{ph}})&=0,\nonumber\\
b^2&=\frac{R_{\text{ph}}^2}{f(R_{\text{ph}})},
\end{align}
where $R_{\text{ph}}$ is the photon-sphere radius and
\begin{equation}\label{eq:lapse}
f(R)\equiv 1-\frac{2\mu r_0}{R}-\frac{\nu r_0^{2\alpha}}{R^{2\alpha}}-\frac{2R}{r_0}.
\end{equation}
\textcolor{black}{In the original generalized Vaidya coordinates this result corresponds to}
\begin{equation}
r_{\rm ph}(v)=\frac{R_{\rm ph}}{r_0}v.
\end{equation}
\textcolor{black}{Thus the physical areal radius of the photon surface evolves with $v$.} 

For a generic value of $\alpha$, the first equation in \eqref{eq:conditions} cannot be solved analytically in a compact form. Therefore, it is useful to apply the perturbative method developed in Ref.~\cite{bib:ali2024podu}. This method relates the shift of the photon-sphere and shadow radii to a deformation of a known reference geometry.

In Ref.~\cite{bib:ali2024podu}, the deformation is usually introduced with respect to the Schwarzschild lapse function. In that case one writes
\begin{equation}
f(R)=\left(1-\frac{2M}{R}\right)e^{g(R)}.
\end{equation}
The photon-sphere radius is then treated as a perturbation around the Schwarzschild value
\begin{equation}
R_{\text{ph}}=3M.
\end{equation}
The sign of $g(R)$ and its derivative at $R=3M$ determines whether the shadow and photon-sphere radii increase or decrease relative to their Schwarzschild values. In particular,
\begin{itemize}
\item if $g(3M)>0$, the shadow radius decreases relative to the Schwarzschild case;
\item if $g(3M)<0$, the shadow radius increases relative to the Schwarzschild case;
\item if $g'(3M)>0$, the photon-sphere radius increases;
\item if $g'(3M)<0$, the photon-sphere radius decreases.
\end{itemize}

However, using the Schwarzschild geometry as the reference background is not the most natural choice for the present spacetime. Indeed, when $\nu=0$, the metric reduces to the Vaidya geometry written in conformally static coordinates $\{t,R,\theta,\varphi\}$. The shadow of the Vaidya spacetime was studied in Ref.~\cite{bib:germany}, while the effect of the charged sector, corresponding to $\alpha=1$ in the present notation, was investigated in Ref.~\cite{bib:heydarzade2024bonnorvaidya}.

For this reason, it is more appropriate to regard the term proportional to $\nu$ as a deformation of the conformally static Vaidya geometry rather than as a deformation of the Schwarzschild metric. We therefore write
\begin{equation}
f(R)=\left(1-\frac{2\mu r_0}{R}-\frac{2R}{r_0}\right)e^{g(R)},
\end{equation}
where
\begin{equation}
\label{eq:g}
g(R)=
\ln\left|
1-\frac{\nu r_0^{2\alpha}R^{1-2\alpha}}
{R-2\mu r_0-\dfrac{2R^2}{r_0}}
\right|.
\end{equation}
This representation isolates the contribution of the Husain matter sector and allows one to analyze how the parameter $\nu$, or equivalently the effective charge-like contribution encoded in $N$, modifies the Vaidya photon sphere and shadow.

Before studying the influence of $\nu$ on the shadow, we first recall the derivation of the photon-sphere radius in the conformally static Vaidya spacetime, following Ref.~\cite{bib:germany}.
\section{Shadow in the Vaidya spacetime admitting a homothetic Killing vector}

The Vaidya solution in advanced Eddington--Finkelstein coordinates~\cite{bib:vaidya} is described by the line element
\begin{equation}\label{eq:usualvaidya}
ds^2=-f(v,r)dv^2+2dvdr+r^2d\Omega^2,
\end{equation}
where
\begin{equation}
f(v,r)=1-\frac{2M(v)}{r},
\end{equation}
and
\begin{equation}
d\Omega^2=d\theta^2+\sin^2\theta\,d\varphi^2
\end{equation}
is the metric on the unit two-sphere. Here $v$ denotes the advanced time coordinate. The spacetime admits a homothetic Killing vector when the mass function is linear in $v$~\cite{bib:nelvaidya},
\begin{equation}
M(v)=\mu v,
\end{equation}
where $\mu>0$ is a constant. \textcolor{black}{This case is especially useful because the homothetic symmetry allows one to reduce the null geodesic equation to a conformally static coordinates. The corresponding constant $R$ photon sphere must, however, be interpreted in the original coordinates as self-similar photon surfaces with $r_{\rm ph}(v)$ proportional to $v$.}

To introduce conformally static coordinates, we perform the transformation $(v,r)\to(t,R)$~\cite{bib:germany},
\begin{align}\label{eq:trans}
v &= r_0 e^{\frac{t}{r_0}},\\
r &= R e^{\frac{t}{r_0}},
\end{align}
where $r_0$ is an arbitrary constant with the dimension of length. Under this transformation the Vaidya metric takes the form
\begin{equation}
ds^2=e^{\frac{2t}{r_0}}d\tilde{s}^2,
\end{equation}
where the conformally related static metric is
\begin{equation}\label{eq:vaidya_conformal}
d\tilde{s}^2
=
-\left(1-\frac{2\mu r_0}{R}-\frac{2R}{r_0}\right)dt^2
+2dtdR+R^2d\Omega^2.
\end{equation}
Since null geodesics are invariant under conformal transformations up to reparametrization, the shadow can be analyzed using the metric $d\tilde{s}^2$. The conformal factor does not change the null trajectories, although it does affect the affine parametrization.

The vector $\partial_t$ is a Killing vector of the conformally related metric $d\tilde{s}^2$ and corresponds to the homothetic Killing vector of the original Vaidya geometry. It is timelike in the region
\begin{equation}
R_-<R<R_+,
\end{equation}
where $R_-$ and $R_+$ are the inner and outer conformal horizons of the conformally static metric. 	They are determined by the zeros of
\begin{equation}
1-\frac{2\mu r_0}{R}-\frac{2R}{r_0}=0
\end{equation}
and are given by
\begin{equation}\label{eq:timelikehkv}
R_{\pm}=\frac{r_0}{4}\left(1\pm\sqrt{1-16\mu}\right).
\end{equation}
\textcolor{black}{These roots should not be confused with the apparent horizon of the original 
Vaidya/Husain spacetime. They are the zeros of the lapse function in the 
conformally static coordinates and therefore define conformal Killing  horizons, or equivalently the boundaries of the region where the homothetic  Killing vector is timelike. Below, when we discuss inner horizon we assume inner conformal Killing horizon and outer horizon we refer as outer conformal Killing horizon. The notions conformal Killing horizon should not be mixed up with the notion of apparent horizon of a black hole. Especially it conserns the outer conformal Killing horizon which is no related to black hole boundary at all.}

Thus, the existence of this static region requires
\begin{equation}
\mu\leq \frac{1}{16}.
\end{equation}

Due to spherical symmetry, photon motion can be restricted to the equatorial plane,
\begin{equation}
\theta=\frac{\pi}{2}.
\end{equation}
The conserved angular momentum is
\begin{equation}
L=R^2\frac{d\varphi}{d\lambda},
\end{equation}
where $\lambda$ is a parameter along the null geodesic. The Killing symmetry of $d\tilde{s}^2$ gives another conserved quantity,
\begin{equation}
E=
\left(1-\frac{2\mu r_0}{R}-\frac{2R}{r_0}\right)\frac{dt}{d\lambda}
-\frac{dR}{d\lambda}.
\end{equation}

Using the null condition,
\begin{equation}
g_{ik}u^iu^k=0,
\end{equation}
one obtains the radial equation
\begin{equation}
\left(\frac{dR}{d\lambda}\right)^2+V_{\text{eff}}=E^2,
\end{equation}
where the effective potential is
\begin{equation}\label{eq:radial_vaidya}
V_{\text{eff}}
=
\left(1-\frac{2\mu r_0}{R}-\frac{2R}{r_0}\right)\frac{L^2}{R^2}.
\end{equation}

Circular photon orbits are determined by
\begin{equation}
V_{\text{eff}}=E^2,\qquad
\frac{dV_{\text{eff}}}{dR}=0.
\end{equation}
Solving these equations gives two possible photon-sphere radii,
\begin{equation}\label{eq:vaidya_condition1}
R_{\text{ph}}^\pm
=
\frac{r_0}{2}
\left(1\pm\sqrt{1-12\mu}\right).
\end{equation}
The reality condition for these radii is
\begin{equation}
\mu\leq \frac{1}{12}.
\end{equation}
However, the existence of the static region already imposes the stronger condition $\mu\leq 1/16$.

\textcolor{black}{In the Vaidya coordinates, the corresponding self-similar photon surface is obtained from $r=Rv/r_0$}:
\begin{equation}
r_{\rm ph}(v) =\frac{v}{2}
\left(1-\sqrt{1-12\mu}\right).
\end{equation}
For $\mu\ll1$, this gives
\begin{equation}
r_{\rm ph}(v)=3\mu v+\mathcal{O}\left(\mu^2\right )=3M(v)+\mathcal{O}\left(\mu^2\right ).
\end{equation}
Thus the usual Schwarzschild relation is recovered only at constant time and only in the slow-accretion limit. 

The corresponding impact parameter is obtained from the condition
\begin{equation}
b_V^2=\frac{L^2}{E^2}.
\end{equation}
Using the photon-sphere condition, one obtains
\begin{equation}\label{eq:vaidya_condition2}
b_V=
\sqrt{
\frac{\left(R_{\text{ph}}^-\right)^3}
{4\mu r_0-R_{\text{ph}}^-}
}.
\end{equation}
For $\mu\ll 1$, this reduces to
\begin{equation}
b_V\approx 3\sqrt{3}\,\mu r_0.
\end{equation}
Again, this is precisely the Schwarzschild value $b_{\text{sh}}=3\sqrt{3}M$ after identifying $M=\mu r_0$ in the conformally static frame.

\textcolor{black}{The observer is located at fixed $R=R_o$ in the conformally static coordinates. In the Vaidya spacetime, this is not a fixed areal radius observer. Instead, the corresponding worldline is}
\begin{equation} \label{eq:observer}
r_o(v)=\frac{R_o}{r_0}v.
\end{equation}
\textcolor{black}{Therefore Eq.~\eqref{eq:vaidya_angular} gives the angular shadow radius
measured by a homothetic observer. The angular radius of the shadow measured by such a homothetic observer is
determined by}

\begin{equation}\label{eq:vaidya_angular}
\sin^2\omega_{\text{sh}}
=
\frac{
b_V^2
\left(
R_o-2\mu r_0-\frac{2R_o^2}{r_0}
\right)
}
{R_o^3}.
\end{equation}
Equivalently,
\begin{equation}
\sin^2\omega_{\text{sh}}
=
\frac{b_V^2}{R_o^2}
\left(
1-\frac{2\mu r_0}{R_o}-\frac{2R_o}{r_0}
\right).
\end{equation}
This form makes clear that the angular size of the shadow is controlled by the lapse function of the conformally static geometry evaluated at the observer's position.

Several limiting cases are important.

First, as the observer approaches either horizon,
\begin{equation}
R_o\to R_\pm,
\end{equation}
the factor
\begin{equation}
1-\frac{2\mu r_0}{R_o}-\frac{2R_o}{r_0}
\end{equation}
vanishes. Therefore,
\begin{equation}
\sin^2\omega_{\text{sh}}\to 0.
\end{equation}
This limit alone does not distinguish between $\omega_{\text{sh}}\to 0$ and $\omega_{\text{sh}}\to \pi$; the physical branch must be fixed by the observer's position relative to the photon sphere.

Near the outer horizon,
\begin{equation}
R_o\to R_+,
\end{equation}
the angular radius tends to zero,
\begin{equation}
\omega_{\text{sh}}\to 0.
\end{equation}
Thus, an observer close to the outer conformal horizon sees a bright sky. In the original coordinates, using
\begin{equation}
r=R\frac{v}{r_0},
\end{equation}
and expanding $R_+$ for $\mu\ll 1$, we obtain
\begin{equation}
R_+\approx \frac{r_0}{2}-2\mu r_0,
\end{equation}
hence
\begin{equation}
r_o\approx \frac{v}{2}-2\mu v
=
\frac{v}{2}-r_{\text{ah}},
\end{equation}
where
\begin{equation}
r_{\text{ah}}=2\mu v
\end{equation}
is the apparent horizon radius of the Vaidya spacetime.

Second, when the observer is located at the photon sphere,
\begin{equation}
R_o=R_{\text{ph}}^-,
\end{equation}
the angular radius satisfies
\begin{equation}
\sin^2\omega_{\text{sh}}=1,
\end{equation}
or
\begin{equation}
\omega_{\text{sh}}=\frac{\pi}{2}.
\end{equation}
Thus, the shadow covers exactly one half of the observer's sky. In the original coordinates and for $\mu\ll 1$, this corresponds to
\begin{equation}
r_o\approx 3\mu v.
\end{equation}

Finally, near the inner horizon,
\begin{equation}
R_o\to R_-,
\end{equation}
the angular radius approaches
\begin{equation}
\omega_{\text{sh}}\to \pi.
\end{equation}
Therefore, an observer close to the inner horizon sees an almost completely dark sky. For small $\mu$,
\begin{equation}
R_-\approx 2\mu r_0,
\end{equation}
and hence, in the original coordinates,
\begin{equation}
r_o\approx 2\mu v=r_{\text{ah}}.
\end{equation}
This shows that the inner horizon of the conformally static metric corresponds, in the original Vaidya coordinates, to the apparent horizon of the radiating black hole.
\section{Shadow in the Husain solution}

We now return to the Husain metric and analyze how the additional matter sector
affects the homothetic photon surface and the shadow measured by homothetic
observers.  Using the definition of the function $g(R)$ given in \eqref{eq:g},
\begin{equation}
g(R) = \ln \left| 1 - \frac{\nu r_0^{2\alpha}R^{1-2\alpha}}{R - 2\mu r_0 - \frac{2R^2}{r_0}} \right|,
\end{equation}
we evaluate it at $R=R_v$, where
\begin{equation}
R_v = \frac{r_0}{2} \left( 1 - \sqrt{1 - 12\mu} \right),
\end{equation}
is the radius of the photon sphere in the Vaidya spacetime.

Since the photon-sphere radius lies in the region of outer communication, the quantity
\begin{equation}
R_v - 2\mu r_0 - \frac{2R_v^2}{r_0} = \frac{r_0}{2} \left( 8\mu - 1 + \sqrt{1 - 12\mu} \right) \ge 0, \quad \mu \in \left[0, \frac{1}{16}\right],
\end{equation}
is positive. Therefore, the sign of $g(R_v)$ is determined by the sign of $\nu$. This leads to two qualitatively different regimes.

For $\alpha>\frac{1}{2}$, the weak energy condition requires $\nu<0$, and hence $g(R_v)>0$. According to the perturbative method described above, this means that the shadow radius of the Husain black hole is smaller than the shadow radius of the corresponding Vaidya black hole.

For $\alpha<\frac{1}{2}$, the weak energy condition requires $\nu>0$, and hence $g(R_v)<0$. In this case, the shadow radius of the Husain black hole is larger than the shadow radius of the corresponding Vaidya black hole.

Thus, within the perturbative regime around the self-similar Vaidya background, the Husain matter sector changes the shadow in a branch-dependent way. 
More precisely, for sufficiently small deformation,
\begin{equation}
\left|
\frac{\nu r_0^{2\alpha}R_v^{\,1-2\alpha}}
{4\mu r_0-R_v}
\right|\ll 1,
\end{equation}
the sign of $g(R_v)$ determines the first-order shift of the critical impact parameter relative to the Vaidya value. Since
\begin{equation}
\delta \ln b_{\rm ph}=-\frac12 g(R_v)+O(\nu^2),
\end{equation}
one finds that
\begin{itemize}
\item for $0\leq \alpha<\frac12$, the weak energy condition requires $\nu>0$, hence $g(R_v)<0$, and the shadow is larger than in the corresponding Vaidya spacetime;
\item for $\frac12<\alpha\leq 1$, the weak energy condition requires $\nu<0$, hence $g(R_v)>0$, and the shadow is smaller than in the corresponding Vaidya spacetime.
\end{itemize}
Therefore, the physically admissible low-pressure branch enlarges the shadow, whereas the physically admissible high-pressure branch reduces it relative to the self-similar Vaidya background.

A completely analogous conclusion holds for the photon sphere. 
Let
\begin{equation}
F(R)\equiv \frac{f'(R)}{f(R)}-\frac{2}{R}.
\end{equation}
The photon sphere is determined by the equation $F(R_{\rm ph})=0$. Writing
\begin{equation}
f(R)=f_V(R)e^{g(R)},
\end{equation}
where $f_V(R)$ is the self-similar Vaidya lapse function, one has
\begin{equation}
F(R)=F_V(R)+g'(R).
\end{equation}
Expanding around the Vaidya photon sphere $R_v$, defined by $F_V(R_v)=0$, gives
\begin{equation}
\delta R_{\rm ph}=-\frac{g'(R_v)}{F_V'(R_v)}.
\end{equation}
Since $F_V'(R_v)<0$ in the physical parameter range, the sign of the photon-sphere shift is the same as the sign of $g'(R_v)$. Using
\begin{equation}
g'(R_v)=
\frac{2(1+\alpha)\nu r_0^{2\alpha}R_v^{-2\alpha}}
{\left(4\mu r_0-R_v\right)
\left(1-\dfrac{\nu r_0^{2\alpha}R_v^{1-2\alpha}}{4\mu r_0-R_v}\right)},
\end{equation}
we conclude that
\begin{itemize}
\item for $0\leq \alpha<\frac12$, one has $\nu>0$ and therefore $\delta R_{\rm ph}>0$;
\item for $\frac12<\alpha\leq 1$, one has $\nu<0$ and therefore $\delta R_{\rm ph}<0$.
\end{itemize}
Hence, within the weak-energy-condition branches, the low-pressure Husain sector enlarges both the photon sphere and the shadow, while the high-pressure branch decreases both relative to the self-similar Vaidya geometry.

Finally, we note a potentially important observational implication. Since the branch $\frac12<\alpha\leq 1$ decreases the shadow relative to the self-similar Vaidya background, it is the phenomenologically more interesting branch when one seeks models that yield smaller apparent shadows than the corresponding uncharged radiating reference geometry.

However, solutions with $\alpha<\frac{1}{2}$ cannot be entirely dismissed. They are disfavored only when one discusses black holes such as M87$^*$ and Sagittarius A$^*$, whose observed shadows are smaller than the Schwarzschild or Vaidya reference values. For other compact objects or different observational regimes, the branch $\alpha<\frac{1}{2}$ may still be physically relevant.

To complement the perturbative analysis, it is useful to consider special values of the equation-of-state parameter for which the photon-sphere equation can be treated exactly or quasi-exactly. These numerical cases provide a direct check of the general branch structure derived above and clarify how the Husain matter sector modifies the optical geometry relative to the self-similar Vaidya background.

\subsection{Exact example: the case \texorpdfstring{$\alpha=0$}{alpha=0}}

The simplest example occurs for
\begin{equation}
\alpha=0,
\end{equation}
for which the equation of state reduces to pressureless matter,
\begin{equation}
P=0.
\end{equation}
In this case the self-similar conformally static lapse function becomes
\begin{equation}
f(R)=1-\frac{2\mu r_0}{R}-\nu-\frac{2R}{r_0}.
\end{equation}
The photon-sphere condition
\begin{equation}
f'(R_{\rm ph})R_{\rm ph}-2f(R_{\rm ph})=0
\end{equation}
can now be solved exactly.

Indeed, since
\begin{equation}
f'(R)=\frac{2\mu r_0}{R^2}-\frac{2}{r_0},
\end{equation}
we obtain
\begin{align}
0&=f'(R_{\rm ph})R_{\rm ph}-2f(R_{\rm ph}) \nonumber\\
&=\left(\frac{2\mu r_0}{R_{\rm ph}^2}-\frac{2}{r_0}\right)R_{\rm ph}
-2\left(1-\frac{2\mu r_0}{R_{\rm ph}}-\nu-\frac{2R_{\rm ph}}{r_0}\right).
\end{align}
After simplification, this gives
\begin{equation}
-2(1-\nu)+\frac{6\mu r_0}{R_{\rm ph}}+\frac{2R_{\rm ph}}{r_0}=0.
\end{equation}
Multiplying by \(R_{\rm ph}r_0/2\), one arrives at the quadratic equation
\begin{equation}
\label{eq:alpha0_quadratic}
R_{\rm ph}^2-(1-\nu)r_0R_{\rm ph}+3\mu r_0^2=0.
\end{equation}
Therefore, the two formal roots are
\begin{equation}
\label{eq:alpha0_roots}
R_{\rm ph}^{\pm}
=
\frac{r_0}{2}
\left[
(1-\nu)\pm\sqrt{(1-\nu)^2-12\mu}
\right].
\end{equation}
The physical unstable photon sphere corresponds to the branch continuously connected to the Vaidya solution when \(\nu\to0\), namely
\begin{equation}
R_{\rm ph}^{(\alpha=0)}
=
\frac{r_0}{2}
\left[
(1-\nu)-\sqrt{(1-\nu)^2-12\mu}
\right].
\end{equation}
The corresponding critical impact parameter is
\begin{equation}
\label{eq:alpha0_bph}
b_{\rm ph}^2=
\frac{\left(R_{\rm ph}^{(\alpha=0)}\right)^2}
{1-\nu-\dfrac{2\mu r_0}{R_{\rm ph}^{(\alpha=0)}}-\dfrac{2R_{\rm ph}^{(\alpha=0)}}{r_0}}.
\end{equation}

Several observations follow immediately.

First, the existence of real circular photon orbits requires
\begin{equation}
\label{eq:alpha0_discriminant}
(1-\nu)^2-12\mu\geq 0.
\end{equation}
Thus, compared with the self-similar Vaidya condition \(\mu\leq 1/12\), the additional Husain parameter \(\nu\) shifts the allowed domain for photon-sphere formation.

Second, for small deformation \(|\nu|\ll1\), expanding Eq.~\eqref{eq:alpha0_roots} around the Vaidya value gives
\begin{equation}
R_{\rm ph}^{(\alpha=0)}
=
R_v+\frac{\nu r_0}{2}
\left(
-1+\frac{1}{\sqrt{1-12\mu}}
\right)
+O(\nu^2),
\end{equation}
where
\begin{equation}
R_v=\frac{r_0}{2}\left(1-\sqrt{1-12\mu}\right)
\end{equation}
is the self-similar Vaidya photon-sphere radius. Since
\begin{equation}
-1+\frac{1}{\sqrt{1-12\mu}}>0
\qquad
\text{for } 0<\mu<\frac{1}{12},
\end{equation}
one finds
\begin{equation}
\nu>0
\quad\Longrightarrow\quad
R_{\rm ph}^{(\alpha=0)}>R_v.
\end{equation}
This agrees exactly with the general perturbative result obtained earlier for the branch \(0\leq \alpha<1/2\): the additional Husain sector enlarges the photon sphere and therefore increases the shadow relative to the self-similar Vaidya geometry.

Finally, the angular shadow radius measured by a static observer located at \(R=R_o\) is
\begin{equation}
\sin^2\omega_{\rm sh}^{(\alpha=0)}
=
\frac{b_{\rm ph}^2}{R_o^2}
\left(
1-\nu-\frac{2\mu r_0}{R_o}-\frac{2R_o}{r_0}
\right).
\end{equation}
Thus, the case \(\alpha=0\) provides an exact example confirming the enlarged-shadow behavior of the weak-energy-condition branch below \(\alpha=1/2\).

\subsection{Exact example: the charged branch \texorpdfstring{$\alpha=1$}{alpha=1}}

The most physically transparent high-pressure branch is obtained for
\begin{equation}
\alpha=1,
\end{equation}
for which
\begin{equation}
P=\rho.
\end{equation}
In this case the additional Husain term scales as \(R^{-2}\), and the conformally static lapse function becomes
\begin{equation}
f(R)=1-\frac{2\mu r_0}{R}-\frac{\nu r_0^2}{R^2}-\frac{2R}{r_0}.
\end{equation}
Since the weak energy condition requires \(\nu<0\) for \(\alpha>1/2\), it is natural to parameterize the additional matter contribution in charge-like form,
\begin{equation}
\nu=-q^2,
\qquad q^2>0,
\end{equation}
so that
\begin{equation}
f(R)=1-\frac{2\mu r_0}{R}+\frac{q^2 r_0^2}{R^2}-\frac{2R}{r_0}.
\end{equation}
This is the self-similar analogue of a charged Vaidya-type geometry.

The photon-sphere condition again reads
\begin{equation}
f'(R_{\rm ph})R_{\rm ph}-2f(R_{\rm ph})=0.
\end{equation}
Now
\begin{equation}
f'(R)=\frac{2\mu r_0}{R^2}-\frac{2q^2 r_0^2}{R^3}-\frac{2}{r_0},
\end{equation}
hence
\begin{align}
0&=f'(R_{\rm ph})R_{\rm ph}-2f(R_{\rm ph}) \nonumber\\
&=
\left(
\frac{2\mu r_0}{R_{\rm ph}^2}
-\frac{2q^2 r_0^2}{R_{\rm ph}^3}
-\frac{2}{r_0}
\right)R_{\rm ph}
\\&-2\left(
1-\frac{2\mu r_0}{R_{\rm ph}}
+\frac{q^2 r_0^2}{R_{\rm ph}^2}
-\frac{2R_{\rm ph}}{r_0}
\right).
\end{align}
After simplification, one finds
\begin{equation}
-2+\frac{6\mu r_0}{R_{\rm ph}}-\frac{4q^2 r_0^2}{R_{\rm ph}^2}+\frac{2R_{\rm ph}}{r_0}=0.
\end{equation}
Multiplying by \(R_{\rm ph}^2 r_0/2\), the photon-sphere equation becomes
\begin{equation}
\label{eq:alpha1_cubic_q}
R_{\rm ph}^3-r_0R_{\rm ph}^2+3\mu r_0^2R_{\rm ph}-2q^2r_0^3=0.
\end{equation}
Equivalently, in terms of the original parameter \(\nu\), one may write
\begin{equation}
\label{eq:alpha1_cubic_nu}
R_{\rm ph}^3-r_0R_{\rm ph}^2+3\mu r_0^2R_{\rm ph}+2\nu r_0^3=0.
\end{equation}
This cubic equation provides the exact example for the branch \(\alpha=1\).

Although the general solution of Eq.~\eqref{eq:alpha1_cubic_q} can be written in Cardano form, it is more instructive to analyze it perturbatively around the Vaidya root. Let
\begin{equation}
R_{\rm ph}=R_v+\delta R,
\end{equation}
where \(R_v\) satisfies
\begin{equation}
R_v^2-r_0R_v+3\mu r_0^2=0.
\end{equation}
Substituting into Eq.~\eqref{eq:alpha1_cubic_q} and retaining only first order in \(q^2\), one obtains
\begin{equation}
\delta R
=
\frac{2q^2r_0^3}{3R_v^2-2r_0R_v+3\mu r_0^2}.
\end{equation}
Using the Vaidya quadratic relation to simplify the denominator,
\begin{equation}
3R_v^2-2r_0R_v+3\mu r_0^2
=
R_v(2R_v-r_0),
\end{equation}
and since for the physical branch \(R_v<r_0/2\), it follows that
\begin{equation}
2R_v-r_0<0.
\end{equation}
Therefore
\begin{equation}
\delta R<0.
\end{equation}
Thus, for the physically admissible charged branch \(\nu<0\), the photon sphere moves inward:
\begin{equation}
R_{\rm ph}^{(\alpha=1)}<R_v.
\end{equation}
This exactly matches the general branch analysis for \(\alpha>1/2\).

The corresponding critical impact parameter is
\begin{equation}
b_{\rm ph}^2
=
\frac{R_{\rm ph}^2}
{1-\dfrac{2\mu r_0}{R_{\rm ph}}+\dfrac{q^2 r_0^2}{R_{\rm ph}^2}-\dfrac{2R_{\rm ph}}{r_0}},
\end{equation}
and the shadow angle seen by a {\color{black} homothetic observer at fixed conformally static coordinate \(R=R_o\)} is
\begin{equation}
\sin^2\omega_{\rm sh}^{(\alpha=1)}
=
\frac{b_{\rm ph}^2}{R_o^2}
\left(
1-\frac{2\mu r_0}{R_o}+\frac{q^2 r_0^2}{R_o^2}-\frac{2R_o}{r_0}
\right).
\end{equation}

Hence, the case \(\alpha=1\) provides an exact charge-like example showing that the weak-energy-condition branch above \(\alpha=1/2\) decreases both the photon-sphere radius and the shadow relative to the self-similar Vaidya background.

\textcolor{black}{For \(\alpha>1/2\), the Husain contribution has the same sign as a charge-like repulsive term, although only the special case \(\alpha=1\) has the standard \(r^{-2}\) charged scaling.}

The two exact examples considered above validate the general perturbative branch structure derived earlier. The pressureless branch \(\alpha=0\), for which the additional matter contribution is positive, enlarges the photon sphere and the shadow. By contrast, the charge-like branch \(\alpha=1\), for which the weak energy condition requires a negative additional contribution, shifts the photon sphere inward and decreases the shadow. These exact results provide a useful bridge between the qualitative sign analysis and the fully dynamical discussion of the next section.

\section{Dynamical evolution of the photon sphere and shadow of Husain black holes}

We now return to the genuinely time-dependent Husain geometry,
\begin{equation}
\label{eq:husain_dyn}
ds^2=-\left(1-\frac{2M(v)}{r}-\frac{N(v)}{r^{2\alpha}}\right)dv^2+2dvdr+r^2d\Omega^2,
\end{equation}
where both $M(v)$ and $N(v)$ are allowed to evolve with the advanced time. 
In this case there is no exact timelike Killing vector, and the shadow is no longer determined by a strictly conserved impact parameter. 
In the following we restrict ourselves to the \textcolor{black}{quasistatic} slow-evolution 
approximation. This means that the characteristic timescale of variation of 
$M(v)$ and $N(v)$ is assumed to be much longer than the orbital timescale of  photons near the unstable null orbit. Under this assumption, one may define an instantaneous photon surface and an  instantaneous critical impact parameter. These quantities should not be  understood as exact conserved structures of the fully dynamical spacetime; they  are local \textcolor{black}{quasistatic} approximations tracking the slow drift of the optical  geometry.

To streamline the analysis, let us introduce the effective mass function
\begin{equation}
\mathcal{M}(v,r)\equiv M(v)+\frac12 N(v)r^{1-2\alpha},
\end{equation}
so that
\begin{equation}
f(v,r)=1-\frac{2\mathcal{M}(v,r)}{r}
=1-\frac{2M(v)}{r}-\frac{N(v)}{r^{2\alpha}}.
\end{equation}
Its $v$-derivative is
\begin{equation}
\partial_v\mathcal{M}(v,r)=\dot M(v)+\frac12 \dot N(v)\,r^{1-2\alpha}
=\dot M(v)+\frac{\dot N(v)}{2r^{2\alpha-1}}.
\end{equation}
The corresponding null-radiation density is proportional to this quantity:
\begin{equation}
\mu_{(n)}(v,r)=\frac{2\,\partial_v\mathcal{M}(v,r)}{r^2}
=\frac{2\dot M(v)}{r^2}+\frac{\dot N(v)}{r^{2\alpha+1}}.
\end{equation}
Therefore, the null energy condition is, in general, \emph{radius dependent}. 
This already shows that one cannot classify the shadow dynamics solely by the signs of $\dot M$ and $\dot N$; the relevant quantity is the local effective influx in the region probed by the unstable null geodesics.

\subsection*{\textcolor{black}{Quasistatic} photon surface}

In the \textcolor{black}{quasistatic} slow-evolution approximation, the instantaneous photon 
surface $r_{\rm ph}(v)$ is defined by the circular-null-orbit condition applied 
to the metric with $v$ fixed:
\begin{equation}
r\,\partial_r f(v,r)-2f(v,r)=0.
\end{equation}
Using
\begin{equation}
f(v,r)=1-\frac{2M(v)}{r}-\frac{N(v)}{r^{2\alpha}},
\end{equation}
one finds
\begin{equation}
\label{eq:inst_photon_eq}
-2+\frac{6M(v)}{r_{\rm ph}}+\frac{2(1+\alpha)N(v)}{r_{\rm ph}^{2\alpha}}=0.
\end{equation}
Equation~\eqref{eq:inst_photon_eq} is the instantaneous photon-surface condition 
within the \textcolor{black}{quasistatic} approximation. It is not an exact circular-orbit condition 
for the fully dynamical spacetime, but the leading-order condition obtained by  constant $M(v)$ and $N(v)$ at fixed $v$.

Differentiating Eq.~\eqref{eq:inst_photon_eq} with respect to $v$ along the trajectory $r=r_{\rm ph}(v)$ gives
\begin{equation}
\label{eq:rphdot_raw}
0=
\frac{6\dot M}{r_{\rm ph}}
+\frac{2(1+\alpha)\dot N}{r_{\rm ph}^{2\alpha}}
+
\left(
-\frac{6M}{r_{\rm ph}^2}
-\frac{4\alpha(1+\alpha)N}{r_{\rm ph}^{2\alpha+1}}
\right)\dot r_{\rm ph}.
\end{equation}
Hence the \textcolor{black}{quasistatic} drift of the photon sphere is
\begin{equation}
\label{eq:rphdot}
\dot r_{\rm ph}
=
\frac{
\dfrac{6\dot M}{r_{\rm ph}}
+
\dfrac{2(1+\alpha)\dot N}{r_{\rm ph}^{2\alpha}}
}{
\dfrac{6M}{r_{\rm ph}^2}
+
\dfrac{4\alpha(1+\alpha)N}{r_{\rm ph}^{2\alpha+1}}
}.
\end{equation}
This equation shows explicitly that the evolution of the photon sphere is controlled by a competition between the mass-accretion rate and the variation of the Husain matter sector.

\subsection*{Instantaneous shadow radius}

At each advanced time $v$, the instantaneous critical impact parameter is
\begin{equation}
b_{\rm ph}^2(v)=\frac{r_{\rm ph}^2(v)}{f\bigl(v,r_{\rm ph}(v)\bigr)}.
\end{equation}
Differentiating this expression and using the photon-sphere relation
\begin{equation}
\partial_r f\bigl(v,r_{\rm ph}\bigr)=\frac{2f\bigl(v,r_{\rm ph}\bigr)}{r_{\rm ph}},
\end{equation}
one obtains a remarkably simple result:
\begin{equation}
\label{eq:bdot_over_b}
\frac{\dot b_{\rm ph}}{b_{\rm ph}}
=
-\frac{1}{2}
\frac{\partial_v f\bigl(v,r_{\rm ph}\bigr)}
{f\bigl(v,r_{\rm ph}\bigr)}.
\end{equation}
Since
\begin{equation}
\partial_v f(v,r)=-\frac{2\dot M(v)}{r}-\frac{\dot N(v)}{r^{2\alpha}},
\end{equation}
Eq.~\eqref{eq:bdot_over_b} becomes
\begin{equation}
\label{eq:bdot_final}
\frac{\dot b_{\rm ph}}{b_{\rm ph}}
=
\frac{1}{2f\bigl(v,r_{\rm ph}\bigr)}
\left(
\frac{2\dot M(v)}{r_{\rm ph}}
+\frac{\dot N(v)}{r_{\rm ph}^{2\alpha}}
\right).
\end{equation}
Because $f\bigl(v,r_{\rm ph}\bigr)>0$ for the unstable null orbit outside the horizon, the sign of $\dot b_{\rm ph}$ is determined by
\begin{equation}
\label{eq:shadow_growth_condition}
2\dot M(v)\,r_{\rm ph}^{2\alpha-1}(v)+\dot N(v).
\end{equation}
Therefore,
\begin{align}
\dot b_{\rm ph}>0
\quad &\Longleftrightarrow \quad
2\dot M\,r_{\rm ph}^{2\alpha-1}+\dot N>0,\\
\dot b_{\rm ph}<0
\quad &\Longleftrightarrow \quad
2\dot M\,r_{\rm ph}^{2\alpha-1}+\dot N<0.
\end{align}
This is the appropriate local criterion for the growth or contraction of the instantaneous shadow in the \textcolor{black}{quasistatic} Husain geometry.

\subsection*{Energy-condition radius}

The hypersurface at which the effective null influx changes sign is determined by
\begin{equation}
\partial_v\mathcal{M}(v,r)=0,
\end{equation}
that is,
\begin{equation}
\dot M(v)+\frac{\dot N(v)}{2r^{2\alpha-1}}=0.
\end{equation}
Whenever the right-hand side is positive and real, this defines the energy-condition radius
\begin{equation}
\label{eq:rnec_correct}
r_{\rm nec}(v)=
\left(
-\frac{\dot N(v)}{2\dot M(v)}
\right)^{\frac{1}{2\alpha-1}}.
\end{equation}
This radius separates regions with positive and negative effective null influx. 
Its physical relevance to the shadow depends on whether it lies inside or outside the photon sphere and the apparent horizon. 
Accordingly, the shadow dynamics is governed not by the global signs of $\dot M$ and $\dot N$ alone, but by the ordering of the three characteristic radii
\begin{equation}
r_{\rm ah}(v),\qquad r_{\rm ph}(v),\qquad r_{\rm nec}(v).
\end{equation}

\subsection*{Charged branch: $\alpha>\frac12$}

The branch $\alpha>\frac12$ is especially interesting because the weak energy condition requires $N(v)<0$, and the additional matter contribution can be written in charge-like form,
\begin{equation}
N(v)\equiv -Q^2(v).
\end{equation}
Then
\begin{equation}
\dot N(v)=-2Q(v)\dot Q(v),
\end{equation}
and the effective mass influx becomes
\begin{equation}
\partial_v\mathcal{M}(v,r)=\dot M(v)-\frac{Q(v)\dot Q(v)}{r^{2\alpha-1}}.
\end{equation}
The instantaneous shadow-growth criterion \eqref{eq:shadow_growth_condition} now reads
\begin{equation}
\dot b_{\rm ph}>0
\quad \Longleftrightarrow \quad
\dot M(v)>\frac{Q(v)\dot Q(v)}{r_{\rm ph}^{2\alpha-1}(v)},
\end{equation}
whereas
\begin{equation}
\dot b_{\rm ph}<0
\quad \Longleftrightarrow \quad
\dot M(v)<\frac{Q(v)\dot Q(v)}{r_{\rm ph}^{2\alpha-1}(v)}.
\end{equation}
Thus, even during accretion, the shadow may either grow or shrink depending on whether the increase in mass dominates over the increase in the charge sector, or vice versa. 
In particular, neutralization processes with $\dot Q<0$ favor shadow growth, while sufficiently rapid charge build-up can reduce the shadow even if $\dot M>0$.

This formulation provides a more precise dynamical interpretation than a classification based only on the signs of $\dot M$ and $\dot Q$. 
The decisive quantity is the local effective influx evaluated in the photon region.

\begin{figure}[t]
    \centering
    \includegraphics[width=0.48\textwidth]{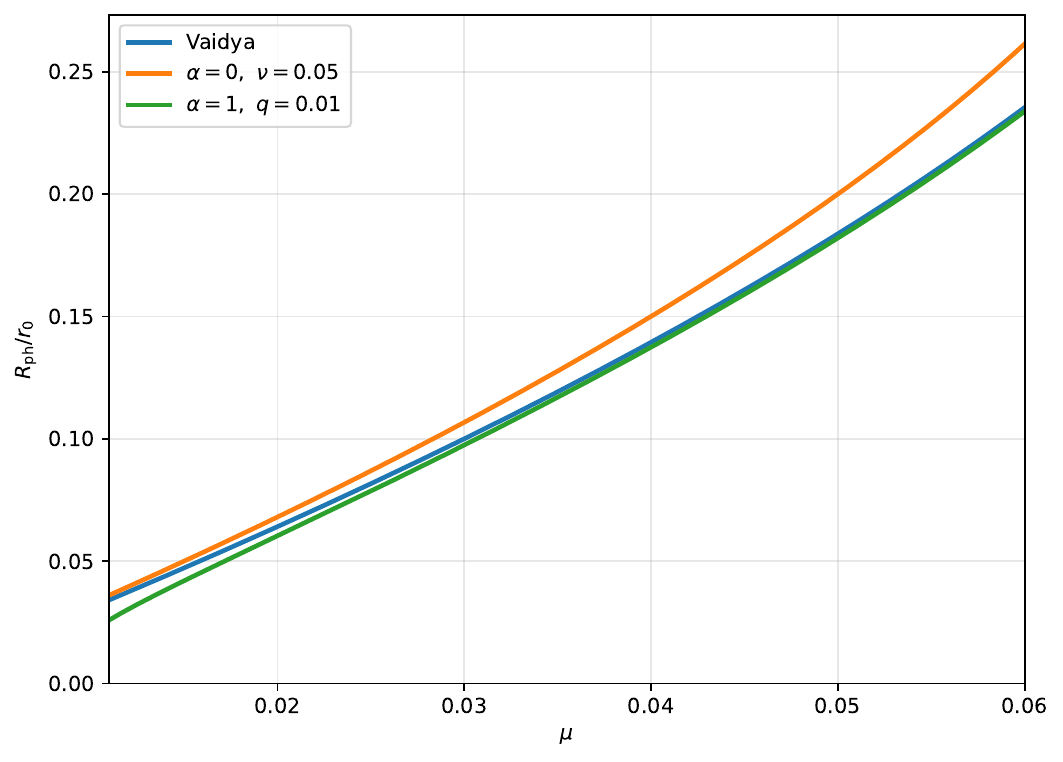}
    \caption{Dimensionless photon-sphere radius $R_{\rm ph}/r_0$ as a function of the self-similar mass parameter $\mu$ for three example cases: the self-similar Vaidya geometry, the exact Husain example $\alpha=0$ with $\nu>0$, and the charge-like example $\alpha=1$ with $\nu=-q^2<0$. The $\alpha=0$ branch lies above the Vaidya curve, showing that the low-pressure weak-energy-condition branch enlarges the photon sphere. By contrast, the $\alpha=1$ branch lies below the Vaidya curve, confirming that the high-pressure charge-like branch shifts the photon sphere inward.}
    \label{fig:Rph}
\end{figure}

Figure~\ref{fig:Rph} illustrates the exact example behavior of the photon-sphere radius in the self-similar Husain geometry. The pressureless branch $\alpha=0$, for which the weak energy condition requires $\nu>0$, produces a photon sphere larger than that of the corresponding self-similar Vaidya spacetime. By contrast, the charge-like branch $\alpha=1$, for which the weak energy condition requires $\nu<0$, yields a smaller photon sphere. The figure therefore confirms, in two exact numerical cases, the general branch structure obtained from the perturbative analysis: the branch $0\leq \alpha<1/2$ enlarges the optical size of the black hole, whereas the branch $1/2<\alpha\leq 1$ decreases it relative to the self-similar Vaidya background.

\section{Conclusions}

\textcolor{black}{In this work, we studied optical properties of the self-similar generalized
Vaidya black hole described by the Husain solution. For the branch admitting a
homothetic Killing vector, we transformed the metric to a conformally static coordinates and analyzed null geodesics by means of the corresponding effective potential. The constant-$R$ photon rings obtained in this metric were then interpreted in the generalized Vaidya coordinates as self-similar photon surfaces}

We showed that the sign of the additional Husain contribution is fixed by the barotropic parameter $\alpha$ once the weak energy condition is imposed. 
Within the perturbative regime around the Vaidya background, the branch $0\leq \alpha<1/2$ enlarges both the photon sphere and the shadow, whereas the branch $1/2<\alpha\leq 1$ reduces them. 
Hence the high-pressure weak-energy-condition branch is the one that naturally leads to smaller shadows than the corresponding Vaidya geometry.

We then considered the genuinely dynamical Husain spacetime. 
In the \textcolor{black}{quasistatic} slow-evolution approximation, we derived an explicit drift 
equation for the instantaneous photon-surface radius and a local criterion for 
the growth or contraction of the critical impact parameter.

The resulting shadow dynamics is governed by the effective influx in the photon region rather than by the signs of $\dot M$ and $\dot N$ alone. 
For the charge-like branch $N=-Q^2$, this yields a transparent condition showing that rapid enough growth of the charge sector may shrink the shadow even during accretion, whereas neutralization tends to enlarge it.

These results clarify the optical role of the Husain matter sector in both the self-similar and slowly evolving regimes, and they provide a useful framework for future phenomenological studies of radiating and charged dynamical black holes.

\acknowledgments

V.V. thanks Eastern Mediterranean University for its hospitality. A. {\"O}. would like to acknowledge the contribution of the COST Action CA21106 - COSMIC WISPers in the Dark Universe: Theory, astrophysics and experiments (CosmicWISPers), the COST Action CA22113 - Fundamental challenges in theoretical physics (THEORY-CHALLENGES), the COST Action CA23130 - Bridging high and low energies in search of quantum gravity (BridgeQG), and the COST Action CA23115 - Relativistic Quantum Information (RQI) funded by COST (European Cooperation in Science and Technology). A. {\"O}. also thank TUBITAK and SCOAP3 for their support.

\bibliography{ref}

\end{document}